\documentclass[preprint,preprintnumbers,amsmath,amssymb,floatfix]{revtex4}
\usepackage{amssymb}
\usepackage[usenames]{color}
\def \blue{}

\begin{document}

\title{ Comment on ``Mass and $K\Lambda$ coupling of N*(1535)''}
\author{S. Ceci, A. \v Svarc{\blue ,} and B. Zauner}
\affiliation{Rudjer Bo\v{s}kovi\'{c} Institute, P.O. Box 180, Bijeni\v{c}ka c. 54, 10 002 Zagreb, Croatia}

\maketitle
It is argued in \cite{Liu96} that when the strong coupling to the $K\Lambda$ channel is considered, {\blue the} Breit-Wigner {\blue (BW)} mass of the lightest orbital excitation of the nucleon N(1535) shifts to {\blue a} lower value. The new value turned out to be smaller than the mass of {\blue the} lightest radial excitation N(1440), which effectively solved the long-standing problem of conventional constituent quark models. In this Comment we show that it is not the Breit-Wigner mass of N(1535) that is decreased, but its bare mass.

According to our understanding, resonance parameters defined by {\blue Eq.} (7) in \cite{Liu96} are much closer to the bare parameters than to the {\blue BW} ones. Therefore, the argumentation given there is in fact related to the general discussion of bare parameter properties which has been going on for quite some time, and was recently summarized in \cite{Bare08}. We present here a general derivation of Eq. (7) in order to clarify the meaning of the resonance mass parameter used in \cite{Liu96}.

A general form of the dressed-propagator denominator from Eq. (7) in \cite{Liu96} is
\begin{equation}\label{eq:denominator}
D(s)=M_0^2-s-\Sigma(s),
\end{equation}
 where $M_0$ is a real-valued mass parameter, $s$ is the Mandelstam variable, and $\Sigma(s)$ is the complex-valued self energy. Since Eq. (7) was a particular extension of the Breit-Wigner approximation ({\em i.e.} Flatt\'e formula \cite{Flatte}) it {\blue may seem} reasonable to call $M_0$ {\blue the} {\blue BW} mass, as in \cite{Liu96}. However, following similar logic, the real part of the pole position \cite{PDG} could also be called {\blue BW} mass, which is not the common practice in N* physics.

Most commonly, the definition of the {\blue BW} mass is based on the resonance peak position. If {\blue the} background is not too strong, the bell-shape of the scattering amplitude will have a peak whenever the denominator of the resonance propagator $D(s)$ is close to its minimum. Unless some exotic form of interaction is considered, and reasonably far from channel openings, the denominator will be minimal when its real part is zero
\begin{equation}\label{eq:s0definition}
  M_0^2-\mathrm{Re}\,\Sigma(s_0)-s_0=0,
\end{equation}
where $s_0$ is {\blue the} value of the physical $s$ for which $\mathrm{Re}\,D(s)=0$.

In \cite{Liu96}, {\blue the} authors cited T\"ornqvist \cite{Tornqvist}, who had defined the {\blue BW} mass $M_{BW}$ as the square root of $s_0$. By this definition, $M_{BW}$ is consistent with the amplitude peak position, which therefore makes it more appropriate to be called {\blue the} Breit-Wigner mass.

Eq. (\ref{eq:denominator}) expressed in terms of $M_{BW}$ instead of $M_0$
\begin{equation}\label{eq:ProperFormula}
D(s)=M_{BW}^2+\mathrm{Re}\,\Sigma(M_{BW}^2)-s-\Sigma(s),
\end{equation}
will help us {\blue understand} under which circumstances $M_0$ may be unambiguously called {\blue BW} mass. This relation looks like Eq. (\ref{eq:denominator}) with an additional term: the real part of $\Sigma(M_{BW}^2)$. {\blue In \cite{Liu96}, the self energy was calculated by Eq. (13) using Flatt\'e formula \cite{Flatte}:
\begin{equation}
\Sigma(s)=i\,M\,\Gamma\,\left(0.8 \rho_{\pi N}(s)+2.1\rho_{\eta N}(s)+3.5\rho_{K\Lambda}(s) \right),
\end{equation}
where $M$ and $\Gamma$ are the N(1535) Breit-Wigner parameters \cite{PDG}, $\rho(s)$ is the phase-space factor for a given channel, and the numerical coefficients are coupling constants. The $K\Lambda$ phase-space factor was then continued below the $K\Lambda$ threshold, giving the explicit imaginary contribution, which effectively generated the mass shift reported in \cite{Liu96}.}


{\blue From Eq. (\ref{eq:ProperFormula}) it is evident that there are two distinct scenarios} when {\blue the} Flatt\'e formula is used. If the resonance peak ({\em i.e.} $M_{BW}$) is above all thresholds, {\blue all phase-space factors will be real and $\mathrm{Re}\,\Sigma(M_{BW}^2)$ will be zero.} In {\blue this} case, calling $M_0$ {\blue the} {\blue BW} mass is perfectly justified. On the other hand, if {\blue $M_{BW}$} is below at least one of the thresholds, {\blue $\mathrm{Re}\,\Sigma(M_{BW}^2)$} will be finite and $M_0$ will no longer be the same as {\blue $M_{BW}$}. {\blue In \cite{Liu96}, the peak of N(1535) was below the $K\Lambda$ channel opening. Therefore the Eq. (\ref{eq:ProperFormula}) should have been used instead of
\begin{equation}
D(s)=M_{BW}^2-s-\Sigma(s),
\end{equation}
to adequately handle the real-valued shift produced by the subthreshold continuation of $\Sigma(s)$.}

To conclude, it was not the Breit-Wigner mass $M_{BW}$ that had to be shifted in order to keep the observed position of {\blue the} N(1535) peak {\blue in} place, but the bare mass $M_0$. If the bare mass turns out to carry some physical meaning \cite{Cec08}, the main result from \cite{Liu96} might have considerable implications. Nevertheless, the precise physical meaning of bare mass is still not fully understood \cite{Bare08}.

\end{document}